\RequirePackage{fix-cm}
\documentclass[smallextended]{svjour3}       % onecolumn (second format)
\smartqed  % flush right qed marks, e.g. at end of proof
\usepackage{graphicx}
\usepackage{graphicx}
\usepackage{amssymb}
\usepackage{amsmath}
\usepackage{bm}
\newcommand{\sgn}{\text{sign}}

\newcommand{\hH}{\text{$\hat{H}$}}
\newcommand{\hD}{\text{$\hat{D}$}}
\newcommand{\hHp}{\text{$\mathcal{\hat{H}}_{\kl}(\gamma)$}}
\newcommand{\rt}{\text{$P_{1/2}$}}
\newcommand{\kl}{\text{$\scriptscriptstyle{<}$}}
\newcommand{\gt}{\text{$\scriptscriptstyle{>}$}}
\newcommand{\sgr}{\text{$\{$sign(x)$\}^{1/2}$}}

\begin{document}

\title{Quantum epistemology from mimicry \& ambiguity.%\thanks{Grants or other notes
}


\author{Han Geurdes         %\and
        %etc.
}


\institute{Han Geurdes \at
              GDS \\
              \email{han.geurdes@gmail.com}           %  \\
}

\date{Received: date / Accepted: date}

\maketitle

\begin{abstract}
In the paper we investigate the epistemological status of quantum theory. 
The starting point is the previously established mathematical ambiguity.
The perspective of our study is the way that Schr{\"o}dinger described Einstein's idea of physics epistemology. 
Namely, physical theory is a map with flags. 
Each flag must, according to Einstein in Schr{\"o}dinger's representation, correspond to a physical reality.
With the ambiguity transformed to operators we were able to mimic certain aspects of quantum theory.
Therefore we claim to have created little flags from the mathematical ambiguity to be put on the map of physics theory.
The question is raised whether nature itself is ambiguous and the mathematical ambiguity is a reflection of it in our language. Or, nature is not ambiguous \& the ambiguity and therefore the necessary Everettian choice of mathematical unversae, can be repaired in mathematics.  
\end{abstract}

\section{Introduction}
\label{intro}
In the paper we will explore the structure of quantum theory to learn the basis of this science in relation to the "peculiarities of the language" used; mathematics. 
Because physics is an old and established natural science, we claim that what we say here will also affect chemistry, biology and fields of (social) science where quantum physics is used \cite{Khren} or is a paradigm. 

Everett solved the mystery of entanglement by introducing, briefly put, many unversae \cite{Ever2}. 
In our paper we will translate many universae into "many points of view" in mathematical unversae.
The physics heresy of Everett \cite{Ever} is transalated to a mathematical heresy in order to find paths were ambiguity in mathematics no longer show the way.

Let us start with some very basic mathematics.
Nobody can doubt the arithmetic fact that, with the real number system at hand, $1+1=(1/2)+(3/2)$. 
Furthermore nobody can doubt the arithmetic fact that $3\times (1/2) =(1/2)\times 3$. 
It also is an elementary arithmetic fact that therefore $y^{1+1}=y^{1/2}y^{3/2}$ and that $y^{3/2}=\{\{y\}^3\}^{1/2}$ and $y^{3/2}=\{\{y\}^{1/2}\}^{3}$ because $3\times (1/2) =(1/2)\times 3$.
To proceed to quantum mechanics. In that theory use is made of complex number wave functions that finally lead to real numbers in Born's probability interpretation.
Therefore we can make no objections to computations where complex numbers occur as an intermediate result as long as the final result is in the real numbers.

As a response to the discussion, described in e.g. \cite{H1}, revolving around Einstein's criticism on the completeness of quantum mechanics \cite{1} Bell wrote an important paper \cite{2}.
In short, Bell supposed that a distinction can be made between entanglement caused by classical hidden variables $\lambda$ and the quantum mechanical description of entanglement. His basic correlation formula embraces all possible hidden variables models.
In very general terms we have.
\begin{eqnarray}\label{B1}
E(a,b)=\int_{\lambda \in \Lambda} \rho(\lambda)A(a,\lambda)B(b,\lambda) d\lambda
\end{eqnarray}
In this formula, $\Lambda$ is the set of values for the $\lambda$. The $a$ and $b$ represent the unit length parameter vectors to measure $\pm 1$ spin. The $\rho(\lambda)$ represents the distribution of the hidden variables $\lambda$. 
The latter is positive definite and normalized, $\int_{\lambda \in \Lambda} \rho(\lambda)=1$. The $A(a,\lambda)$ and $B(b,\lambda)$ represent the measurements of the spin on two distant spin measuring instruments $A$ and $B$. 
Bell, following EPR \cite{1} postulated local hidden variables $\lambda$ \& assumed that $\lambda$ influences the outcome of measurement. Ideally, $A(a,\lambda)=\pm 1$ and $B(b,\lambda)=\pm 1$.

\subsection{The anti-axiom ambiguity}
In recent work we argued that Bell's formula is the source of concrete mathematical incompleteness. 
This work is supported by a complete counter model that can be found in \cite{GKTA}.
The incompleteness of Bell's formula is demonstrated by connecting each possible valid physics model under the umbrella of Bell (\ref{B1}) to the \emph{anti-axiom}; see e.g. \cite[page 10]{Volp}. Below a short version of the derivation of the \emph{anti-axiom} for each Bell model, is presented.

Suppose we define a probability density for a single real variable $x\in \mathbb{R}$. 
\begin{eqnarray}\label{W1}
\rho_{X}(x)=
\left\{
\begin{array}{ll}
-x,~~x\in[-1,0]\\
+x,~~ x\in[0,1] \\
0,~ otherwise
\end{array}\right.
\end{eqnarray}
Obviously, $\rho_{X}(x) \geq 0$ for all $x\in \mathbb{R}$ and $\int \rho_{X}(x) dx =1 $.
Then let us also define a kind of spin function: $\sgn(x)=1$ when $x\geq 0$ and $\sgn(x)=-1$ when $x<0$. We use, $\theta(x)=1$ for $x\geq 0$ and $\theta(x)=0$ for $x<0$ and, $\sgn(x)=2\theta(x)-1$ \cite{Light} and \cite{Fras}.

Please do subsequently note that Bell's formula can always be transformed with $\rho(\lambda,x)\rightarrow \rho(\lambda)\rho_X(x)$ and e.g. $A$ with $A(a,\lambda,x) \rightarrow A(a,\lambda)\sgn(x)$. 
The $\rho_X(x)$ is defined in (\ref{W1}).
There is no physical reason that may disallow it to happen. 
The reason is that there is no explicit theory behind the Bell formula.
It can therefore not be excluded from experiment that the evaluation of each Bell formula is connected to the evaluation of the integral. 
\begin{eqnarray}\label{M1}
E=\int_{-1}^1 |x| \sgn(x) dx
\end{eqnarray} 
In \cite{HanKo1} and \cite{HanKo2} it was demonstrated that $E$ is ambiguous. 
Key element in the argument is $|x|\sgn(x)=x\,\sgn(x)\times \sgn(x)$ in (\ref{M1}).
In order to process the $\sgn(x)\times \sgn(x)$ we note that e.g.
\begin{eqnarray}\label{M1a}
\sgn(x)\times \sgn(x) = \{\sgn(x) \}^{1/2} \times \{\sgn(x) \}^{3/2} 
\end{eqnarray}
is a possibility with complex numbers as intermediate result. 
There can be absolutely no objections against the use of complex numbers as intermediate result.
Of course do observe, as we already alluded to, that $1+1=2=\frac{1}{2}+\frac{3}{2}$.

In the evaluation of $\{\sgn(x)\}^{3/2}$ we may employ two principles to compute $E$. 
Let us study the following
\begin{eqnarray}\label{W10}
F(x)=\{\sgn(x)\}^{3/2}
\end{eqnarray}
Then, we can write down two principles
\begin{itemize}
\item{
\emph{Principle 1:}
The evaluation of $\{\sgn(x)\}^{3/2}$ of the $F(x)$ of (\ref{W10}) is based on first the power $3$ then the power $1/2$. 
}
\item{
\emph{Principle 2:}
The evaluation of $\{\sgn(x)\}^{3/2}$ of the $F(x)$ of (\ref{W10}) is based on first the power $1/2$ then the power $3$.
}
\end{itemize}
Note, furthermore, that $\sgn(x)$ can also be written as: $\exp\left[3i \pi(1-\theta(x))\right]$. 
From the previous we can derive two conflicting forms for the $E$ in (\ref{M1}). According to principle 1
\begin{eqnarray}\label{E1}
E_1=\int_{-1}^0 x \sqrt{\exp\left[3i \pi(1-\theta(x))\right]}\times\sgr dx +
\\\nonumber
\int_0^1 x \sqrt{\exp\left[3i \pi(1-\theta(x))\right]}\times\sgr dx=
\\\nonumber
\int_{-1}^0 x (i\times i) dx +\int_0^1 x dx=1
\end{eqnarray}
and according to principle 2
\begin{eqnarray}\label{E1}
E_2=\int_{-1}^0 x\left( \exp\left[\frac{i \pi}{2}(1-\theta(x))\right]\right)^3\times \sgr dx +
\\\nonumber
\int_0^1  x\left( \exp\left[\frac{i \pi}{2}(1-\theta(x))\right]\right)^3\times \sgr dx=
\\\nonumber
\int_{-1}^0 x ((-i)\times i) dx +\int_0^1 x dx=0
\end{eqnarray}
Therefore, the ambiguity 
\begin{eqnarray}\label{W9}
\left[\left\{\sgn(x)\right\}^3\right]^{1/2}\not\equiv \left[\left\{\sgn(x)\right\}^{1/2}\right]^{3}
\end{eqnarray}
cannot be pushed aside in favour of a preferred particular view concerning Bell type experiments. 

Note that the \emph{anti-axiom} shows from $(1/2) \times 3$  = $3\times (1/2)$.
This implies ambiguity in any physics experiment that follows from Bell's formula and inequalities derived thereof. 
We also note that in defending Bell it is meaningless to fall back on the  computer challenge to reproduce the quantum correlation with a computer simulation of the Bell experiment.
The physics connection of the anti-axiom to the formula destroys its meaning.

\subsection{Epistemology quantum theory}
In a letter from Schr{\"o}dinger to Pauli, Schr{\"o}dinger gives a description of Einstein's view of the epistemology of a physics theory. We quote from \cite{How1}: "He (Einstein) has ....a map with little flags. \emph{To every real thing there must correspond on the map a little flag, and vice versa}".  
Let us employ this nice picture to find out where in quantum theory is the place of the ambiguity.
The question we raise here is related to Wigner's question why mathematics is so effective in describing nature \cite{WigMath}.

In order to perform our study, we try to mimic quantum behaviour with the use of ambiguity based operators. Let us define for $y \in \mathbb{R}$ the operators $\widehat{P_{1/2}}$ and $\widehat{P_3}$
\begin{eqnarray}\label{W20}
\widehat{P_{1/2}}y=\{y\}^{1/2}\\\nonumber
\widehat{P_{3}}y=\{y\}^{3}
\end{eqnarray}
Note, $\widehat{P_{1/2}}(\widehat{P_3}(-1))\not \equiv \widehat{P_{3}}(\widehat{P_{1/2}}(-1))$. The order of the operations is depicted with the brackets.

\section{Quantum mimicry}\label{mim}
We ask the question whether it is possible that epistemological strange characteristics of quantum theory actually arise from ambiguity in mathematics. In order to study the role of the ambiguity in quantum theory, we aim to mimic some aspects of it basing ourselves on the ambiguity. Let us start in the first place to define a wave function for a time-of-observation interval $0\leq t \leq T$ and a space variable $x \in \mathbb{R}$, with $-1\leq x \leq 1$.  
\begin{eqnarray}\label{H1}
\psi(x,t)=e^{it}(\theta(x)-1)C_{\scriptscriptstyle{ <}}+e^{-it}\theta(x)C_{\scriptscriptstyle{>}}
\end{eqnarray}
For the ease of the presentation we repeat the definition already used above. 
The $\theta$ is defined by a Heaviside function
\begin{eqnarray}\label{H1a}
\theta(x)=\left\{
\begin{array}{ll}
1,\hspace{.2 cm} x\geq 0\\
0, \hspace{.2 cm} x<0
\end{array}\right.
\end{eqnarray}
\subsection{Normalization}
The normalization reflects itself in the condition on the constants $C_{\kl}$ and $C_{\gt}$. We have,
\begin{eqnarray}\label{H1b}
|\psi(x,t)|^2= \psi^{*}(x,t)\psi(x,t)= |\psi_{\kl}|^2 + |\psi_{\gt}|^2
\end{eqnarray}
With $\psi_{\kl}=e^{it}C_{kl}(\theta(x)-1)$ and $\psi_{\gt}=e^{-it}C_{>} \theta(x)$.
Equation (\ref{H1b}) holds because obviously from (\ref{H1a}) it follows, $\theta(x)(\theta(x)-1)=0$.
From, (\ref{H1}) we may deduce that ($x<0$)
\begin{eqnarray}\label{H1c}
|\psi_{\kl}|^2=(\theta(x)-1)^2 |C_{\kl}|^2
\end{eqnarray}
and so,
\begin{eqnarray}\label{H1d}
\int_{-1}^1 |\psi_{\kl}|^2 dx =|C_{\kl}|^2 \int_{-1}^0dx=|C_{\kl}|^2
\end{eqnarray}
Furthermore, ($x \geq 0$)
\begin{eqnarray}\label{H1e}
|\psi_{\gt}|^2=\{\theta(x)\}^2 |C_{\gt}|^2=|C_{\gt}|^2
\end{eqnarray}
Therefore,
\begin{eqnarray}\label{H1f}
\int_{-1}^1 |\psi_{\gt}|^2 dx =|C_{\gt}|^2 \int_{0}^1 dx=|C_{\gt}|^2
\end{eqnarray}
and this together with (\ref{H1b}) leads us to
\begin{eqnarray}\label{H1g}
\int_{-1}^1 |\psi(x,t)|^2 dx=|C_{\kl}|^2 + |C_{\gt}|^2=1
\end{eqnarray}
We can conclude that $|C_{\kl}|^2 + |C_{\gt}|^2=1$ is the condition alluded to previously.

\subsection{Hamiltonian}
In the second place, let us define an operator $\hH(x,t)$.
\begin{eqnarray}\label{H2}
\hH(x,t)=ie^{it}C_{\kl}(\theta(x)-1)
\hspace{0.05 cm}
\widehat{\rt} 
\hspace{0.05 cm} 
\widehat{\left( e^{-it}/C_{\kl} \right)}   
+\theta(x)
\end{eqnarray}
With $\hH=\hH_{\kl}+\hH_{\gt}$. In the previous formula especially the wide hat symbols need some attention. The sequence 
\[
\widehat{\rt} 
\hspace{0.05 cm} 
\widehat{\left( e^{-it}/C_{\kl} \right)} f(x,t)
\]
means. First multiply $f(x,t)$ with $e^{-it}/C_{\kl}$. Second, take the square root of that result. In symbols:
\[
\widehat{\rt} 
\hspace{0.05 cm} 
\widehat{\left( e^{-it}/C_{\kl} \right)}f(x,t)
=\{\left( e^{-it}/C_{\kl} \right) f(x,t)\}^{1/2}
=\sqrt{
\left( e^{-it}/C_{\kl} \right) f(x,t)
}
\] 
In a different form with $\widehat{P_3}$ this recipe gives
\[
\widehat{P_3} 
\hspace{0.05 cm} 
\widehat{\left( e^{-it}/C_{\kl} \right)}f(x,t)
=\{
\left( e^{-it}/C_{\kl} \right) f(x,t)\}^3
\]
Let us subsequently suppose that $x<0$. 
Then the outcome of $\hH(x,t)\psi(x,t)$ is
\begin{eqnarray}\label{H3}
\hH(x,t)\psi(x,t)= ie^{it}C_{\kl}(\theta(x)-1)
\hspace{0.05 cm}
\widehat{\rt} 
\hspace{0.05 cm} 
\widehat{\left( e^{-it}/C_{\kl} \right)}
e^{it}C_{\kl}(\theta(x)-1)
\end{eqnarray}
Hence, in $x<0$,
\begin{eqnarray}\label{H4}
\hH(x,t)\psi(x,t)=
ie^{it}C_{\kl}\times(-1)\sqrt{-1}=-\psi(x,t)
\end{eqnarray}
because for $x<0$, we have $\theta(x)-1=-1$ and $\psi(x,t)=e^{it}(-1)C_{\scriptscriptstyle{ <}}\equiv \psi_{\kl}$. 
Furthermore, for $x<0$ we also have
\begin{eqnarray}\label{H5}
i\frac{\partial}{\partial t} \psi(x,t)=-\psi(x,t)
\end{eqnarray}
For $x\geq 0$ it subsequently is easy to acknowledge that $H(x,t)\psi(x,t)=\psi(x,t)$. 
Observe that $\psi_{\gt}=e^{-it}\theta(x)C_{\scriptscriptstyle{>}}$ which is $\psi_{\gt}= e^{-it}C_{\scriptscriptstyle{>}}$ for $x\geq 0$. 
So, $H(x,t)\psi_{\gt}=\psi_{\gt}$.
Moreover, $i\frac{\partial}{\partial t}\psi_{\gt}=\psi_{\gt}$.
Hence, for $t\geq 0$ and $-1\leq x\leq 1$ we may write the Schr{\"o}dinger equation
\begin{eqnarray}\label{H6}
\hH(x,t)\psi(x,t)=i\frac{\partial}{\partial t} \psi(x,t)
\end{eqnarray}
and $\hH(x,t) $ such as defined in (\ref{H3}).
Agreed, the notation is tentative. However a kind of Schr{\"o}dinger equation is derived from an ambiguity producing operator $\widehat{P_{1/2}}$.

\subsection{Hermiticity}
Hermiticity of an operator is an indication for a possible physical existence of the process measured with the operator.
We refer to \cite[page 155]{Merz}. 
Therefore, it is required to have
\begin{eqnarray}\label{H8}
\int_{-1}^1  \psi^{*}(x,t)\hH(x,t) \psi(x,t) dx = \left\{\int_{-1}^1  \psi^{*}(x,t)\hH(x,t) \psi(x,t) dx \right\}^{*}
\end{eqnarray}
For the clarity of presentation let us write $\hH_{\kl}(x,t)$ for $\hH(x,t)$ and $\psi_{\kl}(x,t)$ for $\psi(x,t)$ when $x<0$ 
and $\hH_{\gt}(x,t)$ for $\hH(x,t)$ and $\psi_{\gt}(x,t)$ for $\psi(x,t)$ when $x \geq 0$.
Hence,
\begin{eqnarray}\label{H9}
 \left\{\int_{-1}^1  \psi^{*}(x,t)\hH(x,t) \psi(x,t) dx \right\}^{*}=
 %\\\nonumber
  \left\{\int_{-1}^0  \psi_{\kl}^{*}(x,t)\hH_{\kl}(x,t) \psi_{\kl}(x,t) dx \right\}^{*}
  \\\nonumber +
   \left\{\int_{0}^1  \psi_{\gt}^{*}(x,t)\hH_{\gt}(x,t) \psi_{\gt}(x,t) dx \right\}^{*}
\end{eqnarray}
Looking at the first term on the right hand of (\ref{H9}) we have for $x<0$
\begin{eqnarray}\label{H10}
\left\{\int_{-1}^0  \psi_{\kl}^{*}(x,t)\hH_{\kl}(x,t) \psi_{\kl}(x,t) dx \right\}^{*}=\\\nonumber
\int_{-1}^0 \left( \hH_{\kl}(x,t) \psi_{\kl}(x,t)\right)^{*} \psi_{\kl}(x,t) dx
\end{eqnarray}
Because, 
\[
-\psi_{\kl}^{*}(x,t)=\left( \hH_{\kl}(x,t) \psi_{\kl}(x,t)\right)^{*}=\hH^{*}_{\kl}(x,t)\psi^{*}_{\kl}(x,t)
\]
and so,
\begin{eqnarray}\label{H11}
\left\{\int_{-1}^0  \psi_{\kl}^{*}(x,t)\hH_{\kl}(x,t) \psi_{\kl}(x,t) dx \right\}^{*}=
-\int_{-1}^0|\psi_{kl}(x,t)|^2
\\\nonumber=
\int_{-1}^0  \psi_{\kl}^{*}(x,t)\hH_{\kl}(x,t) \psi_{\kl}(x,t) dx
\end{eqnarray}
Hence in $-1\leq x \leq 0$ the operator is Hermitian.

Looking at the second term on the right hand of (\ref{H9}) we see for $x\geq 0$
\begin{eqnarray}\label{H12}
\left\{\int_{0}^1  \psi_{\gt}^{*}(x,t)\hH_{\gt}(x,t) \psi_{\gt}(x,t) dx \right\}^{*}=
\\\nonumber 
\int_{0}^1  \left(\hH_{\gt}(x,t) \psi_{\gt}(x,t)\right)^{*} \psi_{\gt}(x,t) dx
\end{eqnarray}
Because, \[\psi_{\gt}^{*}(x,t)=\left(\hH_{\gt}(x,t) \psi_{\gt}(x,t)\right)^{*}=\hH_{\gt}^{*}(x,t)\psi_{\gt}^{*}(x,t)\] 
it follows
\begin{eqnarray}\label{H13}
\left\{\int_{0}^1  \psi_{\gt}^{*}(x,t)\hH_{\gt}(x,t) \psi_{\gt}(x,t) dx \right\}^{*}= 
\int_0^1 |\psi_{\gt}(x,t)|^2 dx=
\\\nonumber
\int_{0}^1  \psi_{\gt}^{*}(x,t)\hH_{\gt}(x,t) \psi_{\gt}(x,t) dx
\end{eqnarray}
From the previous equation (\ref{H13}) and equation (\ref{H11}) we can conclude that $\hH(x,t)$ is Hermitian. This implies that a physical reality can exist behind the operator.
\subsection{Commutation} 
\subsubsection{First principle forms}
Let us define an operator $\hD_{\kl}$ for $x<0$, based on the cubic, as
\begin{eqnarray}\label{H14}
\hD_{\kl}= e^{it} (-1)C_{\kl} \widehat{P_3}\widehat{\left( e^{-it}/C_{\kl} \right)}  
\end{eqnarray}
Note, for $x<0$, we have $\psi_{\kl}(x,t)=e^{it}C_{\kl}(\theta(x)-1)$. Hence, 
\begin{eqnarray}\label{H15}
\hD_{\kl}\psi_{\kl}(x,t)=e^{it}C_{\kl}(-1)\{-1\}^3=e^{it}C_{\kl}=-\psi_{\kl}(x,t)
\end{eqnarray}
Let us recall that $x<0$ implies from (\ref{H2}) that
\begin{eqnarray}\label{H16}
\hH_{\kl}=ie^{it}C_{\kl}(-1)\hspace{0.05 cm}
\widehat{
\rt}
\hspace{0.05 cm}
\widehat{\left( e^{-it}/C_{\kl} \right)}
\end{eqnarray}
Therefore we may suspect that  $\left[\hH_{\kl},\hD_{\kl} \right]\not \equiv 0$. 
We note that $\hD_{\kl}$ is Hermitian. 
With $\psi_{\kl}$ and observing (\ref{H15}), an equivalent of equation (\ref{H11}) 
\begin{eqnarray}\label{H16a}
\left\{\int_{-1}^0  \psi_{\kl}^{*}(x,t)\hD_{\kl}(x,t) \psi_{\kl}(x,t) dx \right\}^{*}=
-\int_{-1}^0|\psi_{kl}(x,t)|^2=\\\nonumber
\int_{-1}^0  \psi_{\kl}^{*}(x,t)\hD_{\kl}(x,t) \psi_{\kl}(x,t) dx
\end{eqnarray}
is valid for $\hD_{\kl}$.
\subsubsection{Transformed Hamiltonian form}
Let us subsequently look at a transformed Hamiltonian, $\hHp=\hH_{\kl}+\gamma\hD_{\kl}$.
It is already known that, looking at (\ref{H5}), $\hH_{\kl}\psi_{\kl}=-\psi_{\kl}$ and that $\hD_{\kl}\psi_{\kl}=-\psi_{\kl}$ from (\ref{H15}).
Therefore we arrive at the following eigenvalue equation
\begin{eqnarray}\label{H19}
\hHp \psi_{\kl} =( -1-\gamma)\psi_{\kl}
\end{eqnarray}
Note that the $\widehat{P}$ operators are not neutral for -1. 
Therefore we are allowed to assume additivity but note that $ \phi-\psi_{\kl}=\phi +(-\psi_{\kl})$. We then have for a general form of operator $\hat{O}$ that we want to use here 
\begin{eqnarray}\label{H19x}
\hat{O}(\phi-\psi_{\kl})=\hat{O}\phi +\hat{O}(-\psi_{\kl}).
\end{eqnarray}
The operator $\hHp$ is Hermitian, hence possibly physical, when $\gamma$ is real.
Let us then look at the commutator, $[\hHp,\hD_{\kl}]$.
Firstly looking at (\ref{H14}), (\ref{H16}) and (\ref{H19}) results in:  
\begin{eqnarray}\label{H20}
\hHp\hD_{\kl} \psi_{\kl} = i\psi_{\kl} +\gamma \psi_{\kl}
\end{eqnarray}
If we pay closer attention to the operation $\hH_{\kl}$ in $\hHp$ i.e. (\ref{H16}) the following is seen.
Note $\hD_{\kl}\psi_{\kl}=-\psi_{\kl}$.
We have, looking at  
\begin{eqnarray}\label{H20a}
\hH_{\kl}(-\psi_{\kl})=
ie^{it}C_{\kl}\left(\theta(x)-1\right)
\hspace{0.05 cm}
\widehat{\rt} 
\hspace{0.05 cm} 
\widehat{\left( e^{-it}/C_{\kl} \right)}
e^{it}C_{\kl}=
\\\nonumber
ie^{it}C_{\kl}\left(\theta(x)-1 \right)
\hspace{0.05 cm}
\widehat{\rt} 1=i\psi_{\kl}
\end{eqnarray}
In addition,
\begin{eqnarray}\label{H20b}
\gamma \hD_{\kl}(-\psi_{\kl}) = \gamma e^{it} (-1)C_{\kl} \widehat{P_3}\widehat{\left( e^{-it}/C_{\kl} \right)} e^{it}C_{\kl}=
\\\nonumber
\gamma e^{it} (-1)C_{\kl} \widehat{P_3}1=e^{it} (-1)C_{\kl} 1^3=
\gamma\psi_{\kl}
\end{eqnarray}
Secondly, 
\begin{eqnarray}\label{H21}
\hD_{\kl} \hHp \psi_{\kl} = \hD_{\kl}(-1-\gamma)\psi_{\kl}=\\\nonumber
\hD_{\kl}(-\psi_{\kl}) + \hD_{\kl} (-\gamma \psi_{\kl})=\\\nonumber
\psi_{\kl}+\gamma^3 \psi_{\kl}
\end{eqnarray}
Here we have employed, observing (\ref{H19}), $\hD_{\kl}( -1-\gamma)\psi_{\kl}=\hD_{\kl}(-\psi_{\kl})+\hD_{\kl}(-\gamma\psi_{\kl})$.
The evaluation of the second term is explicitly given by
\begin{eqnarray}\label{H21a}
\hD_{\kl}(-\gamma\psi_{\kl}) = -e^{it}C_{\kl}\widehat{P_3}\widehat{\left( e^{-it} /C_{\kl}\right)}(-\gamma)e^{it}C_{\kl}(-1)=
\\\nonumber
-e^{it}C_{\kl}\widehat{P_3}\gamma=\gamma^3\psi_{\kl}
\end{eqnarray}
This implies that it can be concluded that
\begin{eqnarray}\label{H22}
[\hHp,\hD_{\kl}]\psi_{\kl} = i\psi_{\kl} -(\gamma^3 -\gamma + 1)\psi_{\kl}
\end{eqnarray} 
The previous equation gives $[\hHp,\hD_{\kl}]\psi_{\kl} = i\psi_{\kl}$ and $\gamma^3 -\gamma + 1 =\epsilon $ for $\gamma \approx -1.3247$ with $|\epsilon|\approx 1.0\times 10^{-11}$.
This form is similar to other quantum mechanical non-commutation \cite[page 158, formula 8.66] {Merz}.
We may conclude that equation (\ref{H19}) has, at least approximately, a real eigenvalue $-(1+\gamma)$ and a proper eigen function $\psi_{\kl}(x,t)$ as can be seen from (\ref{H1}). 
This allows a physical equivalent. 
The operator $\hD$ has eigenvalue $-1$ and eigenfunction $\psi_{\kl}$. 
This means that there is a physics possibility as well. 

Now from the commutation $[\hHp,\hD_{\kl}]\psi_{\kl} = i\psi_{\kl}$ we therefore can derive an uncertainty relation noticing that $\psi_{\kl}$ is not normalized but has 
\begin{eqnarray}\label{H22a}
\int_{-1}^1 dx |\psi_{\kl}|^2 =|C_{\kl}|^2
\end{eqnarray}
and $|C_{\kl}|^2$ finite, positive \& real. 
If, similar to the quantum theory we are trying to mimic, we write $\Delta \mathcal{H}_{\kl}$ for the uncertainty in $\hHp$ and $\Delta D_{\kl}$ for uncertainty in $\hD_{\kl}$, we have similar to quantum mechanics \cite[page 158 vv]{Merz}.
\begin{eqnarray}\label{H23}
\Delta \mathcal{H}_{\kl} \Delta D_{\kl} \geq \frac{1}{2}|C_{\kl}|
\end{eqnarray}
The $|C_{\kl}|$ from the previous equation arises from (\ref{H1d}) and (\ref{H22a}).
Obviously, the physical measurements behind $\Delta \mathcal{H}_{\kl}$ and $\Delta D_{\kl}$ are unknown.
But that does not at all mean that those flags on the map refer to nothing.

\subsection{Everettian math universae}
We note that in (\ref{H21}) if $\alpha=(1+\gamma)$ then the outcome will be: 
\begin{eqnarray}\label{H23a}
\hD_{\kl}(( -1-\gamma)\psi_{\kl}) = \hD_{\kl}(-\alpha\psi_{\kl})=\alpha^3\psi_{\kl}
\end{eqnarray}
This does \emph{not} give the $[\hHp,\hD_{\kl}]\psi_{\kl} = i\psi_{\kl}$ for $\gamma \approx -1.3247$.  
Note that $\gamma\approx -2.3247$ with $|\epsilon| \approx 1.823\times 10^{-7}$ again results into $[\hHp,\hD_{\kl}]\psi_{\kl} = i\psi_{\kl}$. 
This is so because $\gamma -(1+\gamma)^3\approx 0$ for $\gamma\approx -2.3247$.

Apparently we have a road to arrive at the commutator with $\gamma \approx -1.3247$ and we have a road that with $\gamma \approx -1.3247$ does not lead to it. 
This looks a bit like an Everettian "many points of view" approach \cite{Ever2} and \cite{Ever} of a \emph{mathematical} universe.  

A critic will of course say that shaky mathematics is saved with an appeal to Everett his view on physics. 
But this critical reader seems to forget that the ambiguity of (\ref{W9}) is behind the "many points of view" of a mathematical universe.
So, in a sense, the consistency of treatment of e.g. Bell's formula such as insisted by a critical reader is no longer a fact. 
It is added that this fact in (\ref{W9}) remains untouched by the rejection or acceptance of papers in journals. 
We also would like to note that it is perhaps a more restricted operator algebra we are looking at when considering (\ref{H19x}). 

\section{Conclusion \& discussion}
The previous sections show that certain aspects of quantum theory can be construed from the mathematical ambiguity in \cite{HanKo1}.
This may shine a different light on what is generally called quantum weirdness.
We set out on a course to search for the ambiguity in uncertainty relations. 
Apparently, we were halted immediately by the necessity of "many points of view" in mathematical unversae in order to find our way. 
Still we found a road from ambiguity to quantum uncertainty.
One may, of course, question the consistency of the road taken. 
But what is consistency when the anti-axiom surfaces.

If we look at what Schr{\"o}dinger wrote about flags on a map representing for Einstein a 1-1 relation between theory and physical entities, it can be observed that the \emph{mathematical ambiguity} in the order of how $\widehat{P}_{1/2}$ and $\widehat{P}_3$ are operated on $(-1)$ i.e.
\begin{eqnarray}\label{H24}
\widehat{P}_{1/2}(\widehat{P}_3(-1))\not \equiv \widehat{P}_{3}(\widehat{P}_{1/2}(-1))
\end{eqnarray}
hides behind the uncertainty relation in (\ref{H23}). 
Note, $\{\{-1\}^3\}^{1/2}=i$ and $\{\{-1\}^{1/2}\}^3=-i$.

The uncertainty for position and momentum measurement \cite{Merz} holds physical reality.
We may, hence, wonder if the flags construed inside the quantum formalism we mimic here, repesent something in the physical reality. 
Let us please observe that the connection of the concepts with the to-be-explained phenomena in the foundation of science is not at all an indisputable certainty.
In terms of the map-and-flags. Nobody knows if flags and reality are connected or perhaps even that placing flags create reality or are introducing a reality on an other map.

Obviously, the finding of the anti-axiom connection to each Bell experiment is a kind of a shock that may, at first sight, appear to have negative consequences only.
Nevertheless, the question we raise here is: is the ambiguity we are looking at an ambiguity of nature or of our language to describe nature?  
The former concurs with Wigner's idea that mathematics effectively describes physical nature.
The concept that the language is the limiting factor concurs with Wittgenstein's view of philosophy \cite[lemma 119]{Witt}. 
There is actually no reason whatsoever why Wittgenstein's lemma 119 might not be true for a theory of natural sciences as well. 

Another point we can raise is the following.
If successful application of mathematics in science is an important trigger for the objectivity of mathematical knowledge \cite{Mol}, then what is the ambiguity and its quantum theoretical application plus Everettian path finding, telling us about mathematical knowledge. 


\end{document}